\documentclass{cmi}
\usepackage{graphicx}
\usepackage{url}
\usepackage{caption}
\usepackage{subcaption}
\usepackage{amssymb}
\usepackage{fancyvrb}
\usepackage{dsfont}
\usepackage{amsmath}
\usepackage{amssymb}
\usepackage{listings}
\usepackage{marvosym}
\usepackage{multirow}
\usepackage{hyperref}
\usepackage{algorithm}
\usepackage[noend]{algpseudocode}
\usepackage{algpseudocode}
\usepackage{colortbl}
\usepackage{float}
\usepackage{subfloat}
\usepackage{tablefootnote}

\makeatletter
\@addtoreset{algorithm}{chapter}
\makeatother

\floatname{algorithm}{Алгоритм}

\DeclareGraphicsExtensions{.pdf}
\graphicspath{{pic/},{pic/pdf/}}

\usepackage{caption}

\begin{document}
	
\classify{УДК 004.272.25, 004.421, 004.032.24}



\newcolumntype{M}[1]{>{\centering\arraybackslash}m{#1}}
\newcolumntype{N}{@{}m{0pt}@{}}

\title{ПАРАЛЛЕЛЬНЫЙ АЛГОРИТМ ПОИСКА ДИССОНАНСОВ \\
ВО ВРЕМЕННОМ РЯДЕ ДЛЯ МНОГОЯДЕРНОГО ПРОЦЕССОРА \\
INTEL XEON PHI KNIGHTS LANDING}

\author{А.В.~Поляков\footnote{Южно-Уральский государственный университет (национальный исследовательский университет), кафедра системного программирования, просп. Ленина, 76, 454080, Челябинск; магистрант, e-mail: \href{mailto:apolyakov@naumen.ru}{apolyakov@naumen.ru}}, М.Л.~Цымблер\footnote{Южно-Уральский государственный университет (национальный исследовательский университет), кафедра системного программирования, просп. Ленина, 76, 454080, Челябинск; к.ф.-м.н., доцент, e-mail: \href{mailto:mzym@susu.ru}{mzym@susu.ru}}}

\maketitle{}

\begin{abstract}
Диссонанс является уточнением понятия аномальной подпоследовательности временного ряда. Задача поиска диссонанса встречается в широком спектре предметных областей, связанных с временными рядами: медицина, экономика, моделирование климата и др. Предложен новый параллельный алгоритм поиска диссонанса во временном ряде для многоядерных систем Intel Xeon Phi поколения Knights Landing (KNL) для случая, когда входные данные могут быть размещены в оперативной памяти. Алгоритм использует возможность независимого вычисления евклидовых расстояний между подпоследовательностями ряда. Распараллеливание выполнено с помощью технологии программирования OpenMP. Алгоритм состоит из двух стадий: подготовка и поиск. На стадии подготовки выполняется построение вспомогательных матричных структур данных, обеспечивающих эффективную векторизацию вычислений на платформе Intel Xeon Phi KNL. На стадии поиска алгоритм находит диссонанс с помощью построенных структур. Представлены результаты вычислительных экспериментов, подтверждающих высокую масштабируемость разработанного алгоритма.
\end{abstract}

\keywords{временной ряд, поиск диссонансов, параллельный алгоритм, OpenMP, Intel Xeon Phi, Knights Landing, векторизация вычислений.}

\markboth{А.В.~Поляков, М.Л.~Цымблер}{Параллельный алгоритм поиска диссонансов временного ряда \ldots}


\section{Введение.}
\label{sec:Introduction}

В настоящее время поиск аномальных подпоследовательностей является одной из наиболее востребованных задач интеллектуального анализа временных рядов в широком спектре практических приложений: моделирование климата, финансовые прогнозы, медицинские исследования и др. 

В работах~\cite{KeoghLF05,DBLP:journals/kais/KeoghLLH07} Кеог (Keogh) и др. предложили термин диссонанс, который является уточнением понятия аномальной подпоследовательности временного ряда. Диссонанс определяется как подпоследовательность ряда, которая имеет максимальное расстояние (минимальную схожесть) до своего ближайшего соседа. Ближайшим соседом подпоследовательности является та подпоследовательность ряда, которая не пересекается с данной и имеет минимальное расстояние до нее (максимальную схожесть) в смысле выбранной меры схожести. 

Понятие диссонанса привлекательно для интеллектуального анализа временных рядов в качестве подхода к нахождению аномалий, поскольку требует только один, интуитивно понятный параметр~--- длина подпоследовательности,~--- в отличие от большинства других алгоритмов поиска аномалий, требующих от трех до семи параметров, не все из которых являются интуитивно понятными~\cite{KeoghLF05}. 

Последовательный алгоритм \emph{HOTSAX}~\cite{KeoghLF05,DBLP:journals/kais/KeoghLLH07} поиска диссонансов использует кодирование временного ряда с помощью техники символьной агрегатной аппроксимации (Symbolic Aggregate ApproXimation, SAX)~\cite{DBLP:conf/dmkd/LinKLC03} и евклидово расстояние в качестве меры схожести. \emph{HOTSAX} итеративно вычисляет евклидовы расстояния между подпоследовательностями ряда и находит максимум среди расстояний до ближайшего соседа. Перебор подпоследовательностей осуществляется в порядке, задаваемом определенной эвристикой, которая позволят отбрасывать подпоследовательности, заведомо не являющиеся диссонансом, без вычисления расстояний.  

В данной статье предлагается новый параллельный алгоритм для поиска диссонансов во временном ряде для многоядерных систем Intel Xeon Phi поколения Knights Landing~\cite{Sodani15} для случая, когда входные данные могут быть размещены в оперативной памяти, получивший название \emph{PhiDD} (Phi Discord Discovery).

Статья организована следующим образом. В разделе~\ref{sec:Definition} приводится формальная постановка задачи и дано краткое описание последовательного алгоритма \emph{HOTSAX}, используемого в качестве базиса нового параллельного алгоритма. Раздел~\ref{sec:Approach} содержит описание алгоритма \emph{PhiDD}. В разделе~\ref{sec:Experiments} описаны вычислительные эксперименты, исследующие эффективность предложенного алгоритма. Заключение резюмирует результаты, полученные в рамках исследования.

\section{Постановка задачи.}
\label{sec:Definition}
 
\subsection{Формальные определения и нотация.}
\label{subsec:fotmalDefinition}

Дадим определения используемых терминов в соответствии с работой~\cite{KeoghLF05}. 

\emph{Временной ряд (time series)} $T$ представляет собой последовательность хронологически упорядоченных вещественных значений: $T=(t_1, t_2, \dots, t_m)$, $t_i \in \mathbb{R}$. Число $m$ обозначается как $\mid T\mid$ и называется длиной ряда.

\emph{Подпоследовательность (subsequence)} $T_{i,\,n}$ временного ряда $T$ представляет собой непрерывное подмножество $T$ из $n$ элементов, начиная с позиции $i$: $T_{i,\,n} = (t_i, t_{i+1}, \dots, t_{i+n-1}), 1 \leqslant n \leqslant m, 1 \leqslant i \leqslant m-n+1$.

Количество подпоследовательностей ряда $T$, имеющих длину $n$, обозначим как $N=m-n+1$.

Подпоследовательности $T_{i,\,n}$ и $T_{j,\,n}$ ряда $T$ называются \emph{непересекающимися (non-self match)}, если $\mid i-j\mid \geqslant n$. Подпоследовательность, которая является непересекающейся к данной подпоследовательности $C$, обозначается как $M_C$.

\emph{Функция расстояния} между подпоследовательностями ряда~--- вещественная функция $Dist: \mathbb{R}^n \times \mathbb{R}^n \rightarrow \mathbb{R}$, которая удовлетворяет аксиомам тождества  и симметрии ($\forall X=T_{i,\,n}, Y=T_{j,\,n}$ $Dist(X,X)=0$ и $Dist(X,Y)=Dist(Y,X)$ соответственно). 

Подпоследовательность $D$ ряда $T$ является \emph{диссонансом (discord)}, если $\forall C, M_C \in T$ $\min(Dist(D,M_D))>\min(Dist(C,M_C))$. Другими словами, некая подпоследовательность ряда является диссонансом, если она имеет максимальное расстояние до ближайшей непересекающейся с ней подпоследовательностью.

\emph{Евклидово расстояние} между двумя подпоследовательностями $X$ и $Y$ длины $n$ временного ряда $T$ вычисляется следующим образом: 

\begin{equation}
	\label{eq:Euclid}
	ED(X,Y) = \sqrt{\sum_{i=1}^{n} (x_i-y_i)^2}  
\end{equation}

\subsection{Последовательный алгоритм.}
\label{subsec:HOTSAX}

Алгоритм \emph{HOTSAX} поиска диссонансов~\cite{KeoghLF05,DBLP:journals/kais/KeoghLLH07} использует евклидову метрику~(\ref{eq:Euclid}) в качестве функции расстояния и предварительную z-нормализацию исходного ряда.

\emph{Z-нормализацией} временного ряда $T$ называется временной ряд $\hat{T}=(\hat{t}_{1},\dots,\hat{t}_{m})$, элементы которого вычисляются следующим образом:

\begin{equation}
	\label{eq:zNormalization}
	\begin{gathered}
		\hat{t_i} = \frac{t_{i}-\mu}{\sigma},\; 1 \leqslant i \leqslant m;~ 
		\mu=\frac{1}{m}\sum\limits_{i=1}^mt_{i};~
		\sigma^2=\frac{1}{m}\sum\limits_{i=1}^mt_i^2-\mu^2.
	\end{gathered}
\end{equation}	

После нормализации среднее арифметическое элементов временного ряда приблизительно равно 0, а среднеквадратичное отклонение близко к 1. 

Подпоследовательности ряда подвергаются \emph{кусочно-агрегирующей аппроксимации (PAA, Piesewise Aggregate Approximation)}~\cite{DBLP:conf/dmkd/LinKLC03}. Кусочно-\-агрегирующим представлением (PAA-представлением) подпоследовательности $C=(c_1,c_2,\dots,c_n)$ является вектор $\overline{C}=(\overline{c}_1,\dots,\overline{c}_w)$, где $w \leqslant n$ и координаты вектора вычисляются следующим образом:

\begin{equation}
	\label{eq:PAA}  
	\overline{c}_i= \frac{w}{n}\cdot\sum_{j=\tfrac{n}{w}\cdot(j-1)+1}^{\tfrac{n}{w} \cdot j} c_j.
\end{equation}

Далее PAA-представление подвергается кодированию с помощью \emph{символьной агрегатной аппроксимации (SAX, Symbolic Aggregate ApproXima\-tion)}~\cite{DBLP:conf/dmkd/LinKLC03}. Символьным представлением (SAX-представлением) подпоследовательности $C=(c_1,c_2,\dots,c_n)$ является \emph{слово} $\hat{C}=(\hat{c}_1,\hat{c}_2,\dots,\hat{c}_w)$, получаемое следующим образом. Пусть имеется символьный алфавит $\mathcal{A}=(\alpha_1,\alpha_2, \dots, \alpha_{\mid\mathcal{A}\mid})$, где запись $\mid\mathcal{A}\mid$ означает мощность алфавита ($\alpha_1=$'a', $\alpha_2=$'b' и т.д.). Тогда 

\begin{equation}
	\label{eq:SAX} 
	\hat{c}_i=\alpha_i \Leftrightarrow \beta_{j-1} \leqslant \hat{c}_i < \beta_j.
\end{equation}

В формуле~(\ref{eq:SAX}) числа $\beta_i$ представляют собой \emph{точки разделения (break\-points)}~\cite{DBLP:conf/dmkd/LinKLC03}, определяемые как упорядоченный список чисел $\mathcal{B}=(\beta_0,\beta_1,\dots,$ $\beta_{\mid\mathcal{A}\mid-1},\beta_{\mid\mathcal{A}\mid})$, где $\beta_0=-\infty$ и $\beta_{\mid\mathcal{A}\mid}=+\infty$, а площадь под кривой нормального распределения $N(0,1)$ между $\beta_i$ и $\beta_{i+1}$ равна $\tfrac{1}{\mid\mathcal{A}\mid}$. Точки разделения для различных значений параметра мощности алфавита могут быть получены из статистических таблиц~\cite{KeoghLF05,DBLP:journals/kais/KeoghLLH07}.

Последовательная реализация поиска диссонансов, предложенная Кеогом и др.~\cite{KeoghLF05,DBLP:journals/kais/KeoghLLH07}, представлена в алг.~\ref{alg:KeoghDD} (здесь запись $S^n_T$ означает множество всех подпоследовательностей ряда $T$, имеющих длину $n$).

\begin{algorithm}[!ht]
	\caption{\textsc{DiscordDiscovery}(\textsc{in} $T,\ n$; \textsc{out} $pos_{bsf}, dist_{bsf}$)}
	\label{alg:KeoghDD}
	\begin{algorithmic}[1]
		\ForAll{$C_i \in S^n_T$}
			\State{$min \leftarrow \infty$}
			\ForAll{$C_j \in S^n_T$ \textbf{and} $|i-j| \geqslant n$}
				\State{$dist \leftarrow \hyperref[eq:Euclid]{\textsc{ED}}(C_i,C_j)$}
				\If{$dist<dist_{bsf}$}
					\State{\textbf{break}}
				\EndIf
				\If{$dist<min$}
					\State{$min \leftarrow dist$}
				\EndIf
			\EndFor
			\If{$min>dist_{bsf}$}
				\State{$dist_{bsf} \leftarrow min$}
				\State{$pos_{bsf} \leftarrow i$}
			\EndIf
		\EndFor
		\State\Return{$\{pos_{bsf}, dist_{bsf}\}$}
	\end{algorithmic}	
\end{algorithm}

\section{Параллельный алгоритм поиска диссонансов \emph{PhiDD}}
\label{sec:Approach}

\subsection{Проектирование алгоритма}
\label{subsec:Design}

Предлагаемый алгоритм использует следующие основные структуры для хранения данных в оперативной памяти.

Исходный временной ряд представляется алгоритмом \emph{PhiDD} в виде матрицы выровненных подпоследовательностей для обеспечения эффективной векторизации вычислений. \emph{Матрица подпоследовательностей} $S_{T}^n \in \mathbb {R}^{N\times(n+pad)}$ определяется следующим образом:

\begin{equation}
	\label{eq:subsequenceMatrix}
	S_{T}^n(i,j):=\tilde{t}_{i+j-1}.
\end{equation} 

Выравнивание подпоследовательностей выполняется следующим образом. Пусть векторный регистр Intel Xeon Phi имеет ширину (т.е. количество вещественных чисел, которые могут быть загружены в векторный регистр за один прием), равную $w$. Если длина искомого диссонанса $n$ не кратна $w$, то каждая подпоследовательность ряда  дополняется фиктивными нулевыми элементами. Обозначим количество фиктивных элементов за $pad = w - (n\;mod\;w)$, тогда выровненная подпоследовательность $\tilde{T}_{i,n}$ определяется следующим образом:

\begin{equation}
	\label{eq:alignedC}
		\tilde{{T}}_{i,n} = \begin{cases}
			t_{i}, t_{i+1},\dots,t_{i+n-1}, \underbrace{0, 0,\dots,0}_{pad}, & if\; n\;mod\;w > 0\\ 
			t_{i}, t_{i+1},\dots,t_{i+n-1}, & otherwise.
		\end{cases}
\end{equation}

Выравнивание данных помогает избежать эффекта разделения цикла (loop peeling)~\cite{Bacon94}, который существенно снижает эффективность векторизации и заключается в следующем. Если начальный адрес массива не выровнен на ширину векторного регистра, то компилятор разбивает цикл на три части, где первая часть итераций, которые обращаются к памяти с начального адреса до первого выровненного адреса, и третья часть итераций с последнего выровненного адреса до конечного адреса, векторизуется отдельно.

Матрица $PAA_T^{n,\,w} \in \mathbb{R}^{N \times w}$ предназначена для хранения \emph{PAA\nobreakdash-предс\-тав\-ле\-ний} подпоследовательностей, полученных в соответствии с (\ref{eq:PAA}).

\emph{Матрица SAX\nobreakdash-представлений}, $SAX_T^{n,\,\mathcal{A}} \in \mathbb{N}^{N \times w}$, хранит символьные подпоследовательности в алфавите $\mathcal{A}$, полученные в соответствии с (\ref{eq:SAX}).

\emph{Индекс потенциальных диссонансов} представляет собой упорядоченный по возрастанию массив $Cand \in \mathbb{N}^N$ с номерами тех подпоследовательностей в матрице подпоследовательностей $S^n_T$, чьи SAX\nobreakdash-представления наиболее редко встречаются в матрице $SAX_T^{n,\,\mathcal{A}}$: 

\begin{equation}
	\label{eq:CandPhiDis2}
	\begin{gathered}
		cand_i=k \Leftrightarrow 
		Freq(k)=\min \limits_{1 \leqslant j \leqslant N}Freq(j) \\
		\forall i<j~~cand_i<cand_j. 
	\end{gathered}
\end{equation}

Здесь $Freq \in \mathbb{N}^N$ представляет собой \emph{частотный индекс}~--- массив, используемый для хранения частот слов из матрицы $SAX_T^{n,\,\mathcal{A}}$: 

\begin{equation}
	\label{eq:FreqPhiDis2}
		freq_i=k \Leftrightarrow 
		\mid\{j: SAX_T^{n,\,\mathcal{A}}(j,\cdot)=SAX_T^{n,\,\mathcal{A}}(i,\cdot) \}\mid=k. 
\end{equation}

Индекс потенциальных диссонансов создается на стадии подготовки данных и далее на стадии поиска задает порядок перебора подпоследовательностей, которые могут являться диссонансами.

\emph{Матрица слов}, $W_{\mathcal{A}} \in \mathbb{N}^{dict_{size} \times w}$, предназначена для хранения всех возможных слов длины $w$, составленных из символов алфавита $\mathcal{A}$. Величина $dict_{size}$ вычисляется как число размещений символов алфавита $\mathcal{A}$ по $w$ символов с повторениями, или  $dict_{size}=\bar{A}^w_{\mid\mathcal{A}\mid}=\mid\mathcal{A}\mid^w$.

В качестве символов алфавита $\mathcal{A}$ будем рассматривать упорядоченный набор натуральных чисел $1, 2, \dots, \mid\mathcal{A}\mid$. Матрица слов организуется таким образом, чтобы все символы каждого слова (элементы в одной строке матрицы) и все слова (строки матрицы) были упорядочены по возрастанию. Для этого вводится хэш-функция $h: \mathbb{N}^w \leftarrow \{1,2,\dots,dict_{size}\}$, определяемая следующим образом:

\begin{equation}
	\label{eq:HashWordIndex}
	h(a_1,a_2, \dots, a_w) =\sum \limits_{j=1}^{w+1}a_j \cdot w^{w-j-1}.
\end{equation}

Значения длины слова и мощности алфавита, $w=4$ и $\mid\mathcal{A}\mid=4$, соответственно, как показывают эксперименты~\cite{KeoghLF05,DBLP:journals/kais/KeoghLLH07}, хорошо подходят при поиске диссонансов во временных рядах из различных предметных областей. В силу этого матрица слов ($4^4 \times 4=256$~элементов) может быть размещена в оперативной памяти.

\emph{Словарный индекс} предназначен для хранения индексов слов алфавита $\mathcal{A}$ в матрице SAX\nobreakdash-представлений и представляет собой матрицу $Index_{Word} \in \mathbb{N}^{dict_{size} \times N}$: 

\begin{equation}
	\label{eq:IndexWordPhiDis2}
	Index_{Word}(i,j)=k \Leftrightarrow 
	W_{\mathcal{A}}(i, \cdot)=SAX_T^{n,\,\mathcal{A}}(k,\cdot).
\end{equation}	

Словарный индекс создается на стадии подготовки данных и далее на стадии поиска используется для задания порядка перебора тех подпоследовательностей, которые не пересекаются с данной. 

\subsection{Реализация алгоритма}
\label{subsec:DesignPrinciples}

Параллельный поиска диссонансов временного ряда \emph{PhiDD} (см. алг.~\ref{alg:PhiDis2}) состоит из двух стадий. На стадии подготовки алгоритм выполняет построение структур данных, рассмотренных выше в разделе~\ref{subsec:Design}. Далее на стадии поиска алгоритм осуществляет нахождение диссонанса с помощью указанных структур. На каждой из этих стадий распараллеливание осуществляется на базе технологии OpenMP~\cite{DBLP:conf/sc/Mattson06}. 

\begin{algorithm}[!ht]
	\caption{\textsc{PhiDD}(\textsc{in} $T, n, w$; \textsc{out} $pos_{bsf}, dist_{bsf}$)}
	\label{alg:PhiDis2}
	\begin{algorithmic}[1]
		\Statex\Comment{Стадия подготовки}
		\State{$S_T^n \leftarrow \hyperref[eq:zNormalization]{\textsc{Znormalize}}(S_T^n)$}
		\State{$W_{\mathcal{A}} \leftarrow \textsc{MakeWordMatrix}(\mathcal{A}, w)$}
		\State{$PAA_T^{n,\,w} \leftarrow \hyperref[eq:PAA]{\textsc{PAA}}(S_T^n,w)$}
		\State{$SAX_T^{n,\,\mathcal{A}} \leftarrow \hyperref[eq:SAX]{\textsc{SAX}}(PAA_T^{n,\,w},\mathcal{A})$}
		\State{$Cand \leftarrow \textsc{MakeCandidates}(SAX_T^{n,\,\mathcal{A}})$}
		\State{$Index_{Word} \leftarrow \textsc{MakeIndexWord}(SAX_T^{n,\,\mathcal{A}})$}
		\Statex\Comment{Стадия поиска}		
		\State{$\{pos_{bsf}, dist_{bsf}\} \leftarrow \hyperref[alg:PotentialDiscord]{\textsc{PotentialDiscord}}(Index_{Word}, Cand, pos_{bsf}, dist_{bsf})$}
		\State{$\{pos_{bsf}, dist_{bsf}\} \leftarrow \hyperref[alg:RefineDiscord]{\textsc{RefineDiscord}}(Index_{Word}, Cand, pos_{bsf}, dist_{bsf})$}
		\State\Return{$\{pos_{bsf}, dist_{bsf}\}$}
	\end{algorithmic}	
\end{algorithm}		

На \emph{стадии подготовки} алгоритм действует следующим образом. Сначала осуществляется построение матрицы подпоследовательностей исходного временного ряда. Далее выполняется генерация матрицы слов, строками которой будут все возможные слова длины $w$, составленные из символов алфавита $\mathcal{A}$. 

После этого происходит формирование PAA\nobreakdash-представления каждой подпоследовательности в матрице подпоследовательностей в соответствии с формулой~(\ref{eq:PAA}). Соответствующий цикл обработки строк матрицы подпоследовательностей распараллеливается с помощью стандартной директивы компилятора \texttt{\#pragma omp parallel for} OpenMP, обеспечивающей статическое разбиение итераций цикла между нитями.

Затем выполняется формирование SAX\nobreakdash-представления подпоследовательности для каждого PAA\nobreakdash-представления, полученного на предыдущем шаге. Параллельная обработка строк матрицы PAA\nobreakdash-представлений, так же как и на предыдущем шаге, обеспечивается использованием директивы компилятора \texttt{\#pragma omp parallel for}.

Далее на основе полученной матрицы SAX\nobreakdash-представлений подпоследовательностей осуществляется формирование индекса потенциальных диссонансов $Cand$ (см. формулу \ref{eq:CandPhiDis2}). Для этого вычисляется частотный индекс $Freq$ (см. формулу \ref{eq:FreqPhiDis2}) и в $Cand$ добавляются номера строк матрицы SAX\nobreakdash-представлений, которые имеют минимальную частоту (встречаются в матрице SAX\nobreakdash-представлений наиболее редко). В данном шаге нахождение минимального элемента массива $Freq$ также распараллеливается с помощью директивы компилятора \texttt{\#pragma omp parallel for} с использованием параметра \texttt{reduction}, который обеспечивает свертку операции поиска минимума.

Следующим шагом алгоритма является построение словарного индекса $Index_{Word}$, выполняемое следующим образом. Осуществляется сканирование матрицы SAX\nobreakdash-представлений. Для каждой строки этой матрицы вычисляется хэш-функция, дающая номер соответствующего элемента в матрице слов (см. формулу \ref{eq:HashWordIndex}). Далее в строку словарного индекса с номером, совпадающим со значением хэш-функции, записывается номер слова в матрице SAX-представлений. В данном шаге цикл обработки строк матрицы SAX\nobreakdash-представлений распараллеливается с помощью стандартной директивы компилятора \texttt{\#pragma omp parallel for} OpenMP, обеспечивающей статическое разбиение итераций цикла между нитями.

\emph{Стадия поиска} состоит из двух следующих шагов. На первом шаге (см. алг.~\ref{alg:PotentialDiscord}) выполняется поиск диссонансов среди подпоследовательностей, входящих в индекс потенциальных диссонансов, построенный на предыдущей стадии, в порядке, задаваемом указанным индексом. Результатом данного шага позиция предполагаемого диссонанса в исходном временном ряде и расстояние (степень схожести) предполагаемого диссонанса с его ближайшим соседом. На втором шаге (см. алг.~\ref{alg:RefineDiscord}) выполняется уточнение найденной ранее позиции предполагаемого диссонанса среди тех подпоследовательностей, которые не входят в в индекс потенциальных диссонансов. 

\begin{algorithm}[!ht]
	\caption{\textsc{PotentialDiscord}(\textsc{in} $T,\ n$; \textsc{out} $pos_{bsf}, dist_{bsf}$)}
	\label{alg:PotentialDiscord}
	\begin{algorithmic}[1]
		\ForAll{$C_i \in Cand$}
		\State{$min \leftarrow \infty$}
		\ForAll{$C_j \in Index_{Word}(SAX_T^{n,\,\mathcal{A}}(C_i))$ \textbf{and} $|i-j| \geqslant n$}
		\State{$dist \leftarrow \hyperref[eq:Euclid]{\textsc{ED}}^2(C_i,C_j)$}
		\If{$dist<dist_{bsf}$}
		\State{\textbf{break}}
		\EndIf
		\If{$dist<min$}
		\State{$min \leftarrow dist$}
		\EndIf
		\EndFor
		\State{\textsf{\#pragma omp parallel for schedule(dynamic)}}
		\ForAll{$C_j \notin Index_{Word}(SAX_T^{n,\,\mathcal{A}}(C_i))$ \textbf{and} $|i-j| \geqslant n$}
		\State{$dist \leftarrow \hyperref[eq:Euclid]{\textsc{ED}}^2(C_i,C_j)$}
		\If{$dist<dist_{bsf}$}	
		\State{\textbf{break}}
		\EndIf		
		\EndFor
		\If{$min>dist_{bsf}$}
		\State{$dist_{bsf} \leftarrow min$}
		\State{$pos_{bsf} \leftarrow i$}
		\EndIf
		\EndFor
		\State\Return{$\{dist_{bsf}, pos_{bsf}\}$}
	\end{algorithmic}	
\end{algorithm}	

Шаг поиска диссонансов (см. алг.~\ref{alg:PotentialDiscord}) заключается в следующем. Для каждой подпоследовательности индекса потенциальных диссонансов $Cand$ выполняется нахождение ближайших соседей среди всех подпоследовательностей временного ряда, в порядке, задаваемом этим индексом, по формуле (\ref{eq:CandPhiDis2}). Из найденных ближайших соседей выбирается сосед, имеющий наименьшую схожесть с другими подпоследовательностями (в смысле Евклидовой метрики). При этом в вычислениях используется квадрат Евклидова расстояния (вместо собственно значения расстояния) для ускорения вычислений.

Поиск ближайшего соседа подпоследовательности осуществляется следующим образом. Сначала в качестве соседей рассматриваются подпоследовательности строки словарного индекса $Index_{Word}$, соответствующей данной обрабатываемой подпоследовательности, которые имеют одинаковое с ней SAX\nobreakdash-представление. Перебор соседей осуществляется в порядке, задаваемом словарным индексом. 

Затем рассматриваются подпоследовательности, не вошедшие в словарный индекс, в порядке, задаваемом матрицей подпоследовательностей. При 
этом расстояние вычисляется только для подпоследовательностей, которые не являются пересекающимися. Если найденное расстояние между подпоследовательностями меньше найденного на предыдущих итерациях цикла расстояния до ближайшего соседа $bsf_{dist}$, то текущая подпоследовательность исключается из рассмотрения. Такое отбрасывание позволяет сократить объем вычислений и возможно, поскольку найдена подпоследовательность, степень схожести которой с текущей подпоследовательностью больше, чем между двумя другими подпоследовательностями, рассмотренными на предыдущей итерации цикла.

После того, как найден ближайший сосед для данной подпоследовательности, производится сравнение его степени схожести со степенью схожести найденного на предыдущих итерациях самого необычного из ближайших соседей. Полученный самый необычный ближайший сосед является предположительным диссонансом, и алгоритм выдает положение диссонанса во временном ряде. 

\begin{algorithm}[!ht]
	\caption{\textsc{RefineDiscord}(\textsc{in} $T,\ n$; \textsc{out} $pos_{bsf}, dist_{bsf}$)}
	\label{alg:RefineDiscord}
	\begin{algorithmic}[1]
		\State{\textsf{\#pragma omp parallel for schedule(dynamic)}}	
		\ForAll{$C_i \notin Cand$}
		\State{$min \leftarrow \infty$}
		\ForAll{$C_j \in Index_{Word}(SAX_T^{n,\,\mathcal{A}}(C_i))$ \textbf{and} $|i-j| \geqslant n$}
		\State{$dist \leftarrow \hyperref[eq:Euclid]{\textsc{ED}}^2(C_i,C_j)$}
		\If{$dist<dist_{bsf}$}
		\State{\textbf{break}}
		\EndIf
		\If{$dist<min$}
		\State{$min \leftarrow dist$}
		\EndIf
		\EndFor
		\ForAll{$C_j \notin Index_{Word}(SAX_T^{n,\,\mathcal{A}}(C_i))$ \textbf{and} $|i-j| \geqslant n$}
		\State{$dist \leftarrow \hyperref[eq:Euclid]{\textsc{ED}}^2(C_i,C_j)$}
		\If{$dist<dist_{bsf}$}
		\State{\textbf{break}}
		\EndIf
		\If{$dist<min$}
		\State{$min \leftarrow dist$}
		\EndIf
		\EndFor		
		\If{$min>dist_{bsf}$}
		\State{$dist_{bsf} \leftarrow min$}
		\State{$pos_{bsf} \leftarrow i$}
		\EndIf
		\EndFor	
		\State\Return{$\{\sqrt{dist_{bsf}}, pos_{bsf}\}$}
	\end{algorithmic}	
\end{algorithm}	

На втором шаге (см. алг.~\ref{alg:RefineDiscord}) выполняется уточнение найденной ранее позиции предполагаемого диссонанса среди тех подпоследовательностей, которые не входят в индекс потенциальных диссонансов. Поиск ближайших соседей каждой из перебираемых подпоследовательностей выполняется аналогично поиску на первом шаге (нахождение предположительного диссонанса, см. алг.~\ref{alg:PotentialDiscord}). Среди всех найденных ближайших соседей осуществляется поиск такого соседа, который имеет наименьшую степень схожести со всеми подпоследовательностями. Результатом данного шага является уточненное положение диссонанса временного ряда и расстояние (степень схожести) от данного диссонанса до его ближайшим соседом.

В обоих описанных выше шагах стадии поиска перебор ближайших соседей распараллеливается с помощью стандартной директивы компилятора OpenMP \texttt{\#pragma omp pa\-ral\-lel for}. Поскольку отбрасывание неперспективных подпоследовательностей приводит неравномерной вычислительной загрузке нитей, в указанной директиве используется параметр \texttt{schedule (dynamic)}, обеспечивающий динамическое распределение итераций цикла между нитями. Операторы тела цикла, выполняющего вычисление квадратов Евклидовых расстояний, векторизуются компилятором.


\section{Вычислительные эксперименты.}
\label{sec:Experiments}
\subsection{Цели, аппаратная платформа и наборы данных экспериментов}
\label{subsec:ExperimentsSetup}

Для исследования эффективности разработанного алгоритма были проведены вычислительные эксперименты. В качестве аппаратной платформы экспериментов использованы вычислительные узлы кластерной системы Сибирского Суперкомпьютерного Центра ИВМиМГ СО РАН~\cite{ClusterIVMiMG}, характеристики которой приведены в \tabref{tab:HardwarePhiDis2}.

\begin{table}[!ht]
	\centering		
	\caption{Аппаратная платформа экспериментов}
	\label{tab:HardwarePhiDis2}
		\begin{tabular}{|l|c|c|c|c|}
			\hline
			Характеристика	& \multicolumn{2}{c|}{Процессор Intel Xeon} & \multicolumn{2}{c|}{Ускоритель Intel Xeon Phi} \\ \hline
			Модель	&    E5-2630v4        &   E5-2697v4        &   SE10X (KNC)        &   7290 (KNL)        \\ \hline
			Кол-во физ. ядер	&   $2\times$10     &    $2\times$16       &     60      &    72       \\ \hline
			Гиперпоточность	&     2$\times$     &      2$\times$    &      4$\times$    &       4$\times$    \\ \hline
			Кол-во лог. ядер	&     40    &     64    &    240    &     288    \\ \hline
			Частота, ГГц	&     2.2   &    2.6    &    1.1    &     1.5    \\ \hline
			Размер VPU, бит	&     256   &    256    &    512    &     512    \\ \hline
			Пиковая произв-ть, TFLOPS	&    0.390  &    0.600  &    1.076  &  3.456         \\ \hline
		\end{tabular}
\end{table}

Исследование производилось на наборах данных, которые представлены в \tabref{tab:DatasetsPhiDis2}. Указанные наборы использовались в экспериментах по исследованию эффективности работы распределенного алгоритма поиска диссонансов \emph{PDD}~\cite{DBLP:conf/pakdd/HuangZMLLWHD16,DBLP:conf/hpcc/WuZHLLL15}.

\begin{table}[!ht]	
	\caption{Наборы данных для экспериментов}
	\label{tab:DatasetsPhiDis2}
	\centering
	\begin{tabular}{|l|c|c|c|}
		\hline
		Набор данных & Вид & $|T|=m$ & $n$ \\ \hline
		SCD-1M &  Реальный & $10^6$ & 64, 128, 512, 1024, 2048 \\ \hline
		SCD-10M & Реальный & $10^7$ & 64, 128, 512, 1024, 2048 \\ \hline
	\end{tabular}
\end{table}

В экспериментах исследовались быстродействие и мас\-шта\-би\-ру\-е\-мость алгоритма \emph{PhiDD}. При исследовании быстродействия учитывалось время работы алгоритма за вычетом времени загрузки данных в память и выдачи результата. Масштабируемость параллельного алгоритма означает его способность адекватно адаптироваться к увеличению параллельно работающих вычислительных элементов (процессов, процессоров, нитей и др.) и характеризуется ускорением и параллельной эффективностью, которые определяются следующим образом~\cite{VoevodinV02}.

\emph{Ускорение} и \emph{параллельная эффективность} параллельного алгоритма, запускаемого на $k$~нитях, вычисляются как $s(k)=\tfrac{t_1}{t_k}$ и $e(k)=\tfrac{s(k)}{k}$ соответственно, где $t_1$ и $t_k$~--- время работы алгоритма на одной и $k$ нитях соответственно.

В экспериментах рассматривались вышеприведенные показатели в зависимости от изменения параметра $n$ (длина искомого диссонанса). 

Время работы алгоритма \emph{PhiDD} сравнивалось с временем работы  оригинального последовательного алгоритма  \emph{HOTSAX}~\cite{DBLP:journals/kais/KeoghLLH07} и распределенным алгоритмом поиска диссонансов \emph{PDD}~\cite{DBLP:conf/pakdd/HuangZMLLWHD16,DBLP:conf/hpcc/WuZHLLL15}.

\subsection{Результаты экспериментов}
\label{subsec:Experiments-Results}

\begin{figure}[!ht]
	\begin{minipage}[t]{0.5\textwidth}
		\includegraphics[width=\linewidth]{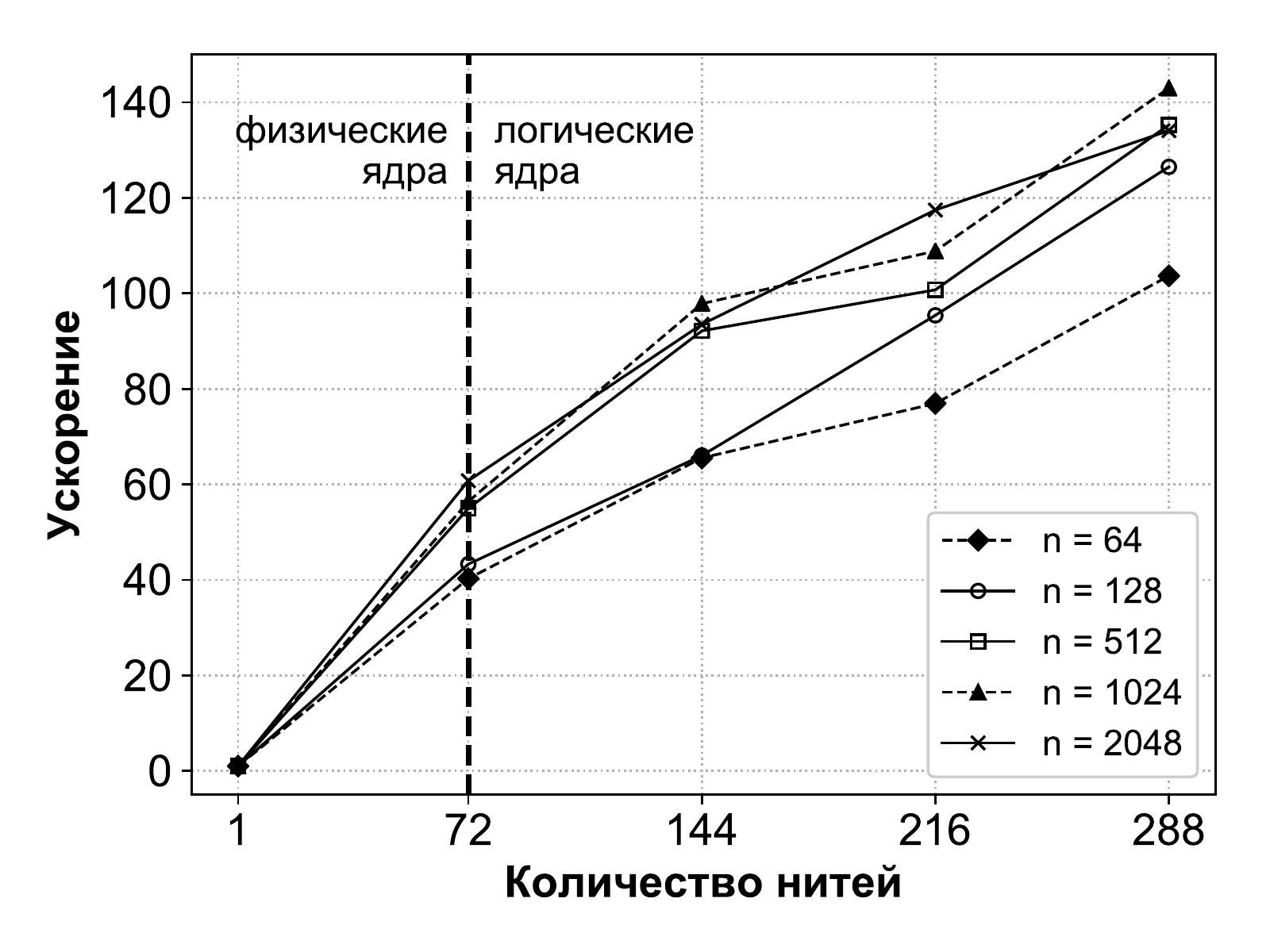}
		\center{а) Ускорение}
		\label{subfig:speedup-SCD1M}
	\end{minipage}
	\hfill
	\begin{minipage}[t]{0.5\textwidth}
		\includegraphics[width=\linewidth]{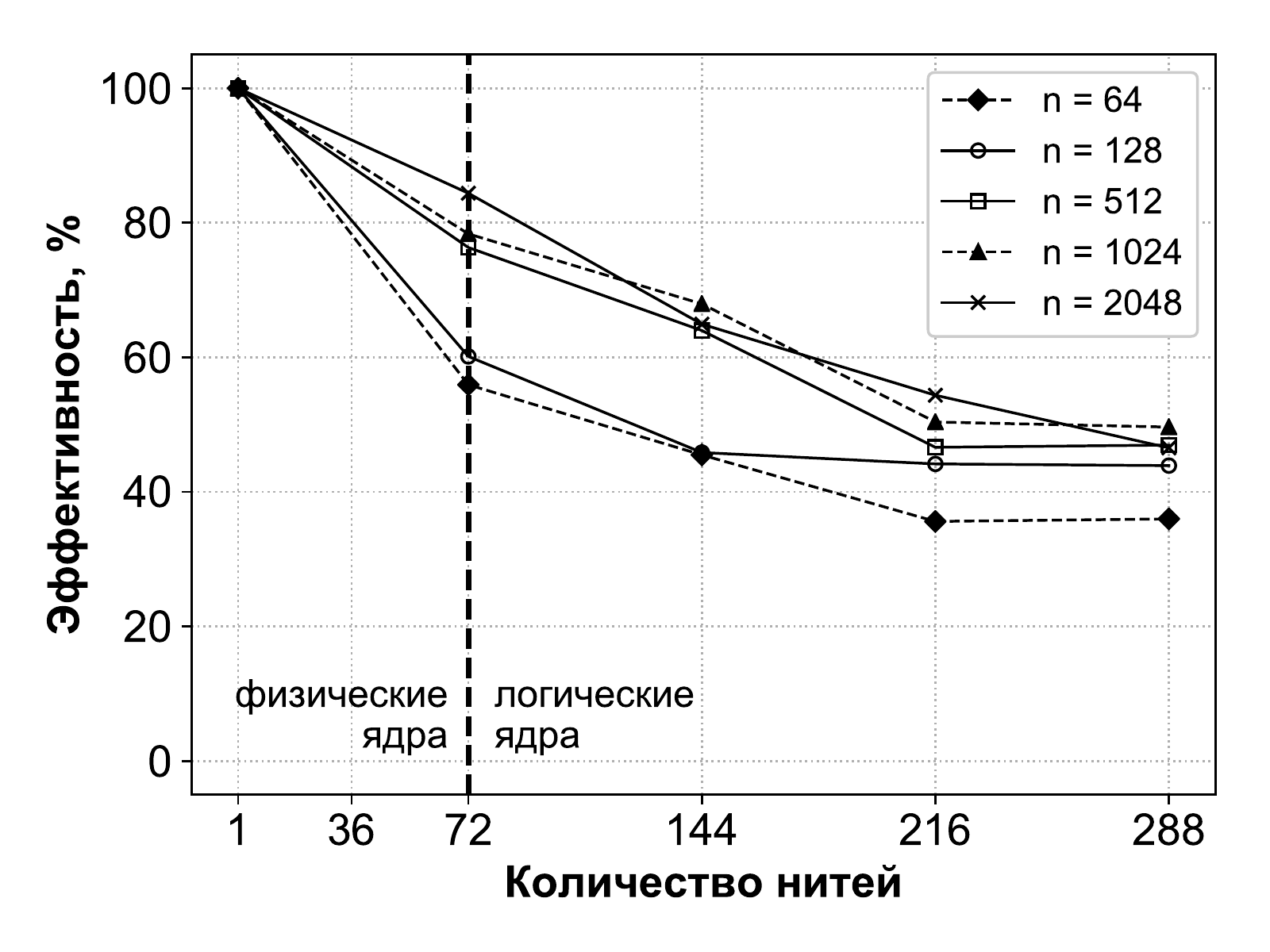}
		\center{б) Параллельная эффективность}
		\label{subfig:efficiency-SCD1M}
	\end{minipage}
	\hspace{0.75em}
	\caption{Масштабируемость \emph{PhiDD} при обработке ряда SCD-1M}
	\label{fig:PhiDis2-Scalability-SCD1M}
\end{figure}

\begin{figure}[!ht]
	\begin{minipage}[t]{0.5\textwidth}
		\includegraphics[width=\linewidth]{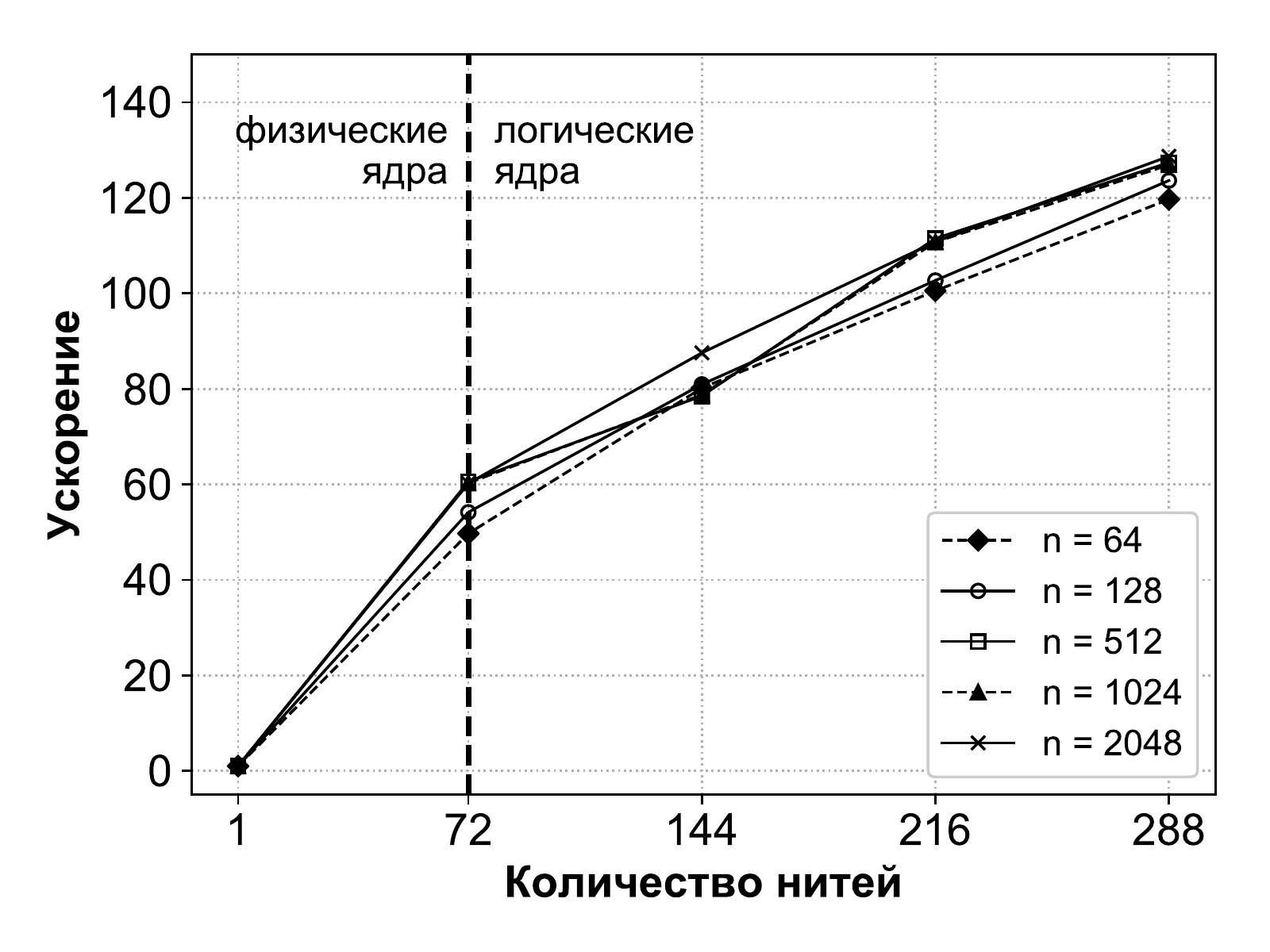}
		\center{а) Ускорение}
		\label{subfig:speedup-SCD10M}
	\end{minipage}
	\hfill
	\begin{minipage}[t]{0.5\textwidth}
		\includegraphics[width=\linewidth]{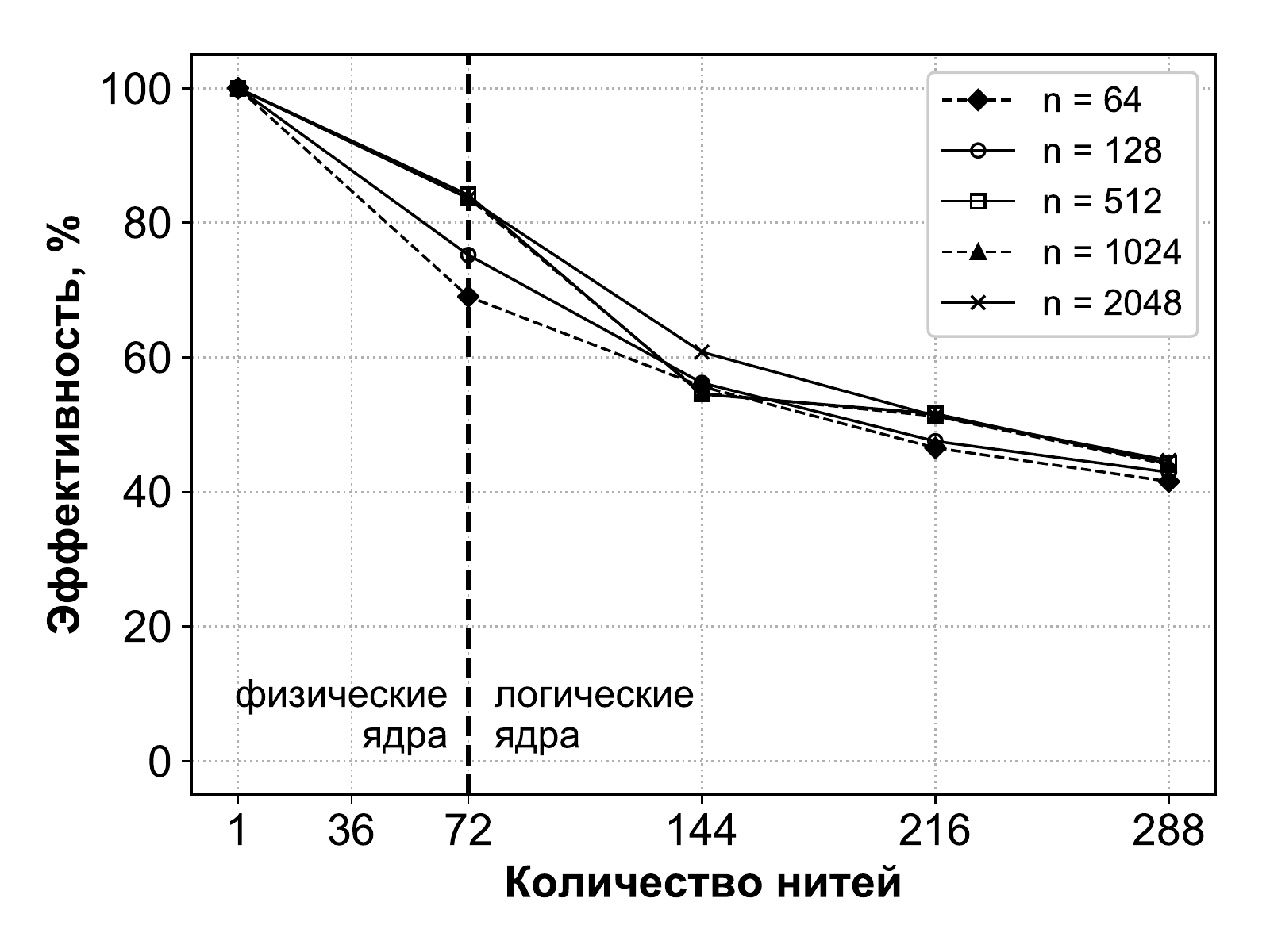}
		\center{б) Параллельная эффективность}
		\label{subfig:efficiency-SCD10M}
	\end{minipage}
	\hspace{0.75em}
	\caption{Масштабируемость \emph{PhiDD} при обработке ряда SCD - 10M}
	\label{fig:PhiDis2-Scalability-SCD10M}
\end{figure}

Результаты экспериментов \emph{по исследованию масштабируемости} алгоритма на системе Intel Xeon Phi Knights Landing представлены на \figref{fig:PhiDis2-Scalability-SCD1M} и \figref{fig:PhiDis2-Scalability-SCD10M}.

Можно видеть, что \emph{PhiDD} демонстрирует ускорение от 40 до 60 и параллельную эффективность от 50 до 90\% (в зависимости от длины искомого диссонанса), если количество нитей, на которых запущен алгоритм, совпадает с количеством физических ядер системы Intel Xeon Phi.

При увеличении количества нитей, запускаемых на одном физическом ядре системы, ускорение является сублинейным, равно как наблюдается и падение параллельной эффективности. При этом наилучшие показатели ускорения и параллельной эффективности ожидаемо наблюдаются при б\'{о}льших значениях параметров $m$ и $n$ (длина исходного временного ряда и длина искомого диссонанса соответственно), обеспечивающих алгоритму наибольшую вычислительную нагрузку. Например, при обработке ряда длиной 1~млн. 288~нитями наблюдаются ускорение~100 и параллельная эффективность 40\% при $n=64$, тогда как при $n=2048$ эти показатели равны 140 и 50\% соответственно. 

Полученные результаты позволяют сделать заключение о хорошей масштабируемости разработанного алгоритма и эффективном использовании им возможностей векторизации вычислений на многоядерной системе Intel Xeon Phi.

\begin{figure}[!ht]
	\begin{minipage}[t]{0.5\textwidth}
		\includegraphics[width=\linewidth]{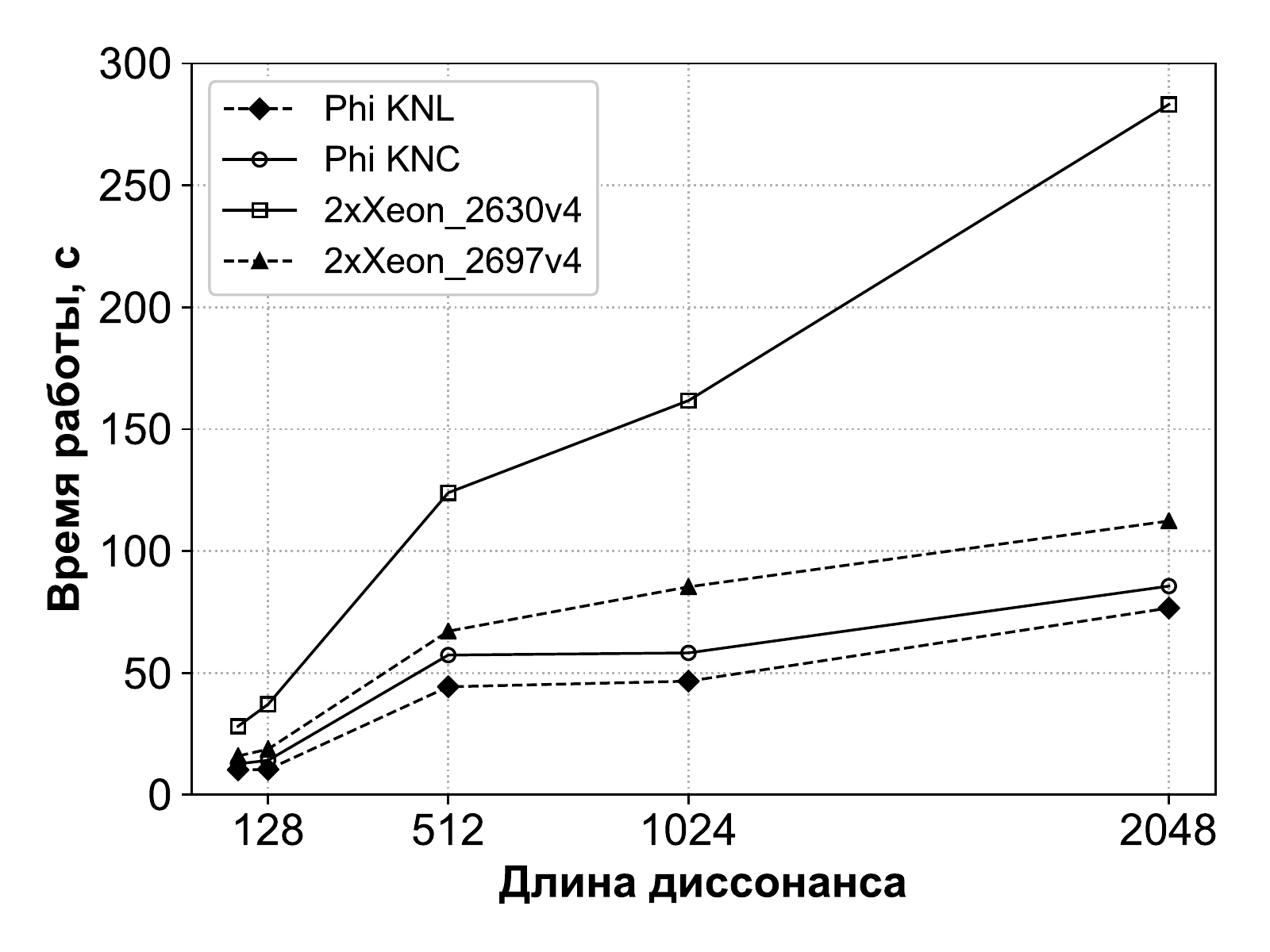}
		\center{а) Набор данных SCD-1M}
		\label{subfig:runtime-SCD1M}
	\end{minipage}
	\hfill
	\begin{minipage}[t]{0.5\textwidth}
		\includegraphics[width=\linewidth]{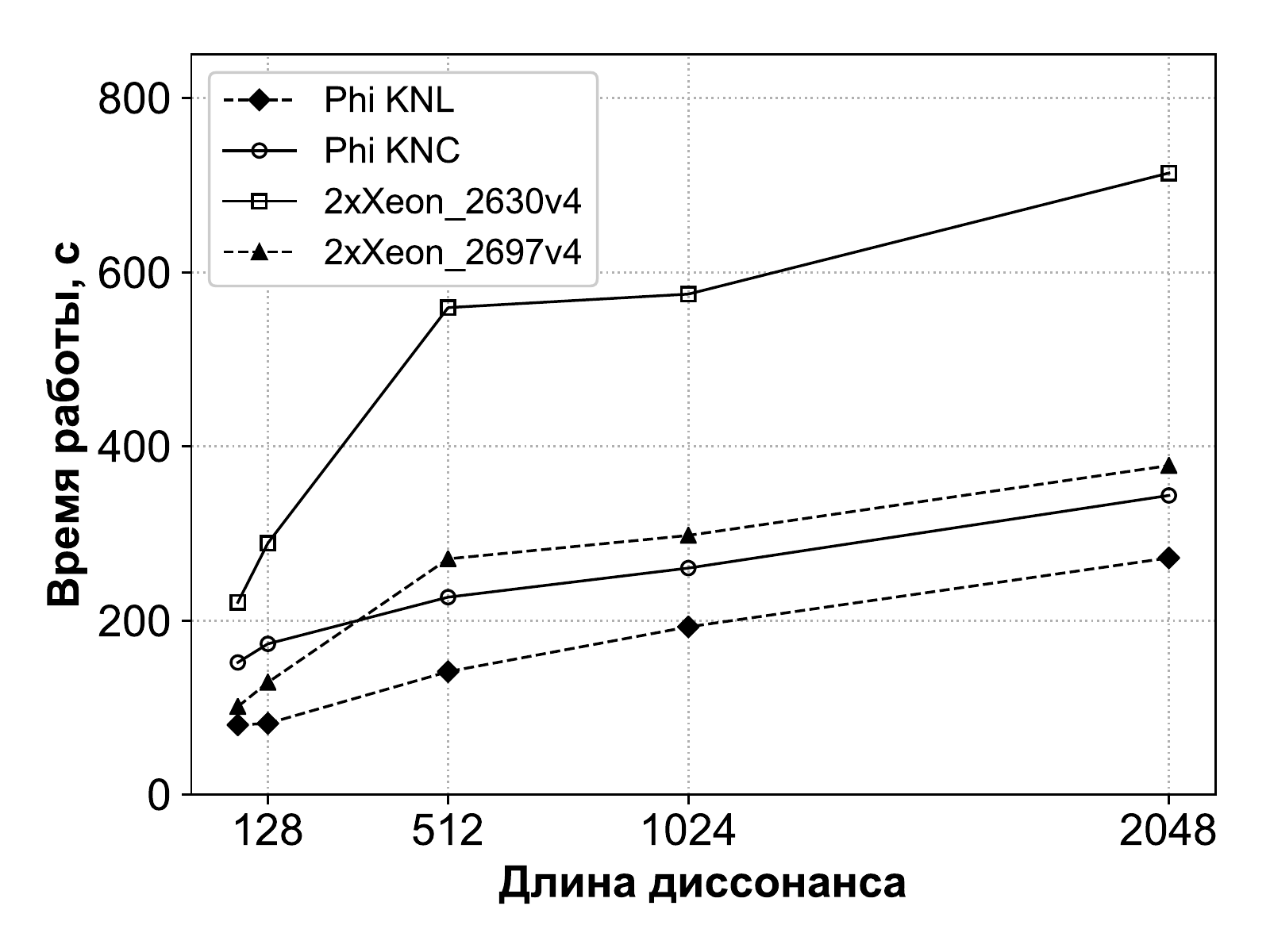}
		\center{б) Набор данных SCD-10M}
		\label{subfig:runtime-SCD10M}
	\end{minipage}
	\hspace{0.75em}
	\caption{Быстродействие алгоритма \emph{PhiDD}}
	\label{fig:PhiDis2-Runtime-SCD}
\end{figure}

Результаты экспериментов \emph{по исследованию быстродействия} алгоритма представлены на рис.~\ref{fig:PhiDis2-Runtime-SCD}. Можно видеть, что алгоритм \emph{PhiDD} работает, как правило, быстрее на платформе многопроцессорной системы Intel Xeon Phi, чем на платформе двухпроцессорного узла Intel Xeon. Данный факт всегда имеет место для более производительного устройства поколения Knights Landing; менее производительное устройство поколения Knights Corner уступает 64\nobreakdash-ядерному двухпроцессорному узлу Intel Xeon при поиске диссонансов длины $n \leqslant 128$ в ряде из 10~млн. точек. В целом на платформе Intel Xeon Phi KNL алгоритм опережает себя на платформе 64\nobreakdash-ядерного двухпроцессорного узла Intel Xeon в 1.5--2~раза.

\begin{table}[!ht]
	\centering
	\caption{Сравнение алгоритма \emph{PhiDD} с аналогами}
	\label{tab:PhiDis2-Comparison}
	\resizebox{\textwidth}{!}{
		\begin{tabular}{|c|c|c|c|c|c|c|c|c|}
			\hline
			\multirow{5}{*}{Набор} & \multicolumn{5}{c|}{Время работы алгоритмов} & \multicolumn{3}{c|}{Ускорение алгоритмов} \\
			& \multicolumn{5}{c|}{на указанных платформах (при $n=128$), с} & \multicolumn{3}{c|}{относительно \emph{HOTSAX}} \\ \cline{2-9} 
			
			&  \emph{PDD}  & \multicolumn{2}{c|}{\emph{PhiDD}} & \multicolumn{2}{c|}{\emph{HOTSAX}} & \multirow{2}{*}{} & \multicolumn{2}{c|}{\multirow{2}{*}{}} \\ \cline{2-6}
			
			& 10~CPU & \multicolumn{2}{c|}{Intel Xeon Phi 7290} & \multicolumn{2}{c|}{Intel Core i5} & \emph{PDD} & \multicolumn{2}{c|}{\emph{PhiDD}} \\ \cline{3-4} \cline{7-9} 
			
			& 1.2~ГГц & 10~нитей &  288~нитей & \multicolumn{2}{c|}{2.6~ГГц} & 10~CPU & 10~нитей & 288~нитей \\ \hline
			
			\multicolumn{1}{|l|}{SCD-1M} & 36~720 & 109.2 & 10.4 & \multicolumn{2}{c|}{147~600} & $4\times$ & $1.3 \cdot 10^3\times$ &  $1.4 \cdot 10^4\times$ \\ \hline
			
			\multicolumn{1}{|l|}{SCD-10M} & 399~600 & 833.3 & 81.9 & \multicolumn{2}{c|}{$\geqslant 432~000$\tablefootnote{Остановлено после 5 суток безрезультатной работы.}} & $\geqslant 1.1$ & $\geqslant 5 \cdot 10^2\times$ & $\geqslant 5 \cdot 10^3\times$ \\ \hline
		\end{tabular}
	}
\end{table}

Результаты экспериментов \emph{по сравнению PhiDD с аналогами} представлены в табл.~\ref{tab:PhiDis2-Comparison}. Данные о быстродействии алгоритма \emph{PDD} взяты из работы~\cite{DBLP:conf/pakdd/HuangZMLLWHD16}. Результаты быстродействия алгоритма \emph{HOTSAX} получены на платформе рабочей станции с процессором Intel Core~i5 2.6~ГГц и использованы для сравнения ускорения, показываемого алгоритмами \emph{PDD} и \emph{PhiDD} относительно \emph{HOTSAX}. Быстродействие \emph{PhiDD} дополнительно показано на 10~ядрах, поскольку в этом случае аппаратная платформа экспериментов имеет примерно равную пиковую производительность с платформой, на которой получены экспериментальные результаты быстродействия алгоритма \emph{PDD}. Можно видеть, что \emph{PhiDD} опережает аналоги.


\section{Заключение.}
\label{sec:Conclusion}
В работе рассмотрена проблема поиска диссонанса во временном ряде. Диссонанс является уточнением понятия аномальной подпоследовательности временного ряда.  Данная задача возникает в широком спектре приложений интеллектуального анализа временных рядов: моделирование климата, финансовые прогнозы, медицинские исследования и др.

Предложен новый параллельный алгоритм поиска диссонанса во временном ряде для многоядерных систем Intel Xeon Phi поколения Knights Landing для случая, когда входные данные могут быть размещены в оперативной памяти, получивший название \emph{PhiDD}. Проведены вычислительные эксперименты, результаты которых показали хорошую масштабируемость алгоритма \emph{PhiDD}.

\vspace{1em}
{Работа выполнена при финансовой поддержке Российского фонда фундаментальных исследований (грант №~17-07-00463), Правительства РФ в соответствии с Постановлением №~211 от 16.03.2013 (соглашение №~02.A03.21.0011) и Министерства образования и науки РФ (государственное задание 2.7905.2017/8.9). Работа выполнена с использованием ресурсов ЦКП Сибирский Суперкомпьютерный Центр ИВМиМГ СО РАН.}

\begin{biblio}

\bibitem{VoevodinV02}
\emph{Воеводин~В.В., Воеводин~Вл.В.} Параллельные вычисления. СПб.:
БХВ-Петербург, 2002.	
	
\bibitem{ClusterIVMiMG}
Перечень оборудования Центра коллективного пользования Сибирского Суперкомпьютерного Центра ИВМиМГ СО РАН. Дата обращения: 03.11.2018. URL: \href{http://www.sscc.icmmg.nsc.ru/hardware.html}{http://www.sscc.icmmg.nsc.ru/hardware.html}.
	
\bibitem{Bacon94}
\emph{Bacon~D.F., Graham~S.L., Sharp~O.J.} Compiler transformations for high-performance computing. ACM Comput. Surv. 1994. \textbf{26}, No.~4. P.~345--420. DOI:~\href{https://dx.doi.org/10.1145/197405.197406}{10.1145/197405.197406}.

\bibitem{DBLP:conf/pakdd/HuangZMLLWHD16}
\textit{Huang~T., Zhu~Y., Mao~Y. et al.} Parallel discord discovery // Proc. of the 20th Pacific-Asia Conference on Advances in Knowledge Discovery and Data Mining, PAKDD 2016, Auckland, New Zealand, April~19--22, 2016. 2016. P.~233--244. DOI:~\href{https://dx.doi.org/10.1007/978-3-319-31750-2_19}{10.1007/978-3-319-31750-2\_19}.

\bibitem{KeoghLF05} 
\textit{Keogh~E., Lin~J., Fu~A.} HOT SAX: efficiently finding the most unusual
time series subsequence // Proc. of the 5th IEEE Int. Conf. on Data Mining, ICDM’05, Houston, Texas, November~27--30,
2005. 2005. P.~8. DOI:~\href{https://dx.doi.org/10.1109/ICDM.2005.79}{10.1109/ICDM.2005.79}.

\bibitem{DBLP:journals/kais/KeoghLLH07} 
\textit{Keogh~E.J., Lin~J., Lee S., Herle~H.V.} Finding the most unusual time
series subsequence: algorithms and applications // Knowl. Inf. Syst.
2007. Vol.~11, No.~1. P.~1--27. DOI:~\href{https://doi.org/10.1007/s10115-006-0034-6}{10.1007/s10115-006-0034-6}.

\bibitem{DBLP:conf/dmkd/LinKLC03}
\textit{Lin~J., Keogh~E.J., Lonardi~S., Chiu~B.Y.} A symbolic representation of
time series, with implications for streaming algorithms // Proc. of
the 8th ACM SIGMOD workshop on Research issues in data mining and
knowledge discovery, DMKD 2003, San Diego, California, USA, June~13,
2003. 2003. P.~2--11. DOI:~\href{https://dx.doi.org/10.1145/882082.882086}{10.1145/882082.882086}.

\bibitem{DBLP:conf/sc/Mattson06}
\textit{Mattson~T.} S08 - Introduction to OpenMP // Proc. of the
ACM/IEEE SC2006 Conf. on High Performance Networking and Computing, November~11--17, 2006, Tampa, FL, USA. 2006. P.~209.
DOI:~\href{https://dx.doi.org/10.1145/1188455.1188673}{10.1145/1188455.1188673}.

%
%
%

\bibitem{Sodani15}
\emph{Sodani~A.} Knights Landing (KNL): 2nd generation Intel Xeon Phi processor // 2015 IEEE Hot Chips 27th Symposium (HCS), Cupertino, CA, USA, August~22--25, 2015. IEEE, 2015. 1--24.

\bibitem{DBLP:conf/hpcc/WuZHLLL15}
\textit{Wu~Y., Zhu~Y., Huang~T. et al.} Distributed discord discovery: Spark
based anomaly detection in time series // Proc. of the 17th IEEE Int. Conf. on High Performance Computing and Communications, HPCC 2015, 7th IEEE Int. Symp. on Cyberspace Safety and Security, CSS 2015, and 12th IEEE Int. Conf. on
Embedded Software and Systems, ICESS 2015, New York, NY, USA, August~24--26, 2015. 2015. P.~154--159. DOI:~\href{https://dx.doi.org/10.1109/HPCC-CSS-ICESS.2015.228}{10.1109/HPCC-CSS-ICESS.2015.228}.

\end{biblio}

\classify{} 

\title{Parallel Algorithm for Time Series Discords Discovery \\
	on the Intel Xeon Phi Knights Landing Many-core Processor}

\author{A.V.~Polyakov$^1$ and M.L.~Zymbler$^2$}

\addressen{%
	$^{1}$ South Ural State University, Department of System Programming; prospekt Lenina 76, Chelyabinsk, 454080, Russia; Master student, e-mail:  \href{mailto:apolyakov@naumen.ru}{apolyakov@naumen.ru}\\
	$^{2}$ South Ural State University, Department of System Programming; prospekt Lenina 76, Chelyabinsk, 454080, Russia; Cand. Sci., Associate Professor, e-mail: \href{mailto:mzym@susu.ru}{mzym@susu.ru}}

\received{}

\maketitle{}

\begin{abstracten}
Discord is a refinement of the concept of anomalous subsequence of a time series. The task of discords discovery is applied in a wide range of subject domains related to time series: medicine, economics, climate modeling, etc. In this paper, we propose a novel parallel algorithm for discords discovery for the Intel Xeon Phi Knights Landing (KNL) many-core systems for the case when input data fit in main memory. The algorithm exploits the ability to independently calculate Euclidean distances between the subsequences of the time series. Computations are paralleled through OpenMP technology. The algorithm consists of two stages, namely precomputations and discovery. At the precomputations stage, we construct the auxiliary matrix data structures, which ensure efficient vectorization of computations on Intel Xeon Phi KNL. At the discovery stage, the algorithm finds discord based upon the structures above. Experimental evaluation confirms the high scalability of the developed algorithm.	

\keywordsen{time series, discords discovery, parallel algorithm, OpenMP, Intel Xeon Phi, Knights Landing, data layout, vectorization.}
\end{abstracten}

\justifying

\begin{biblio_lat}
	
\bibitem{VoevodinV02en}
V.~V.~Voevodin, and Vl.~V.~Voevodin, \emph{The Parallel Computing}. (BHV-Petersburg, St.~Petersburg, 2002).

\bibitem{ClusterIVMiMGen}
Hardware specifications of the Siberian Supercomputing Center. Accessed: 03.11.2018. URL: \href{http://www.sscc.icmmg.nsc.ru/hardware.html}{http://www.sscc.icmmg.nsc.ru/hardware.html}.

\bibitem{Bacon94en}
D.~F.~Bacon, S.~L.~Graham, and O.~J.~Sharp, ``Compiler transformations for high-performance computing,'' ACM Comput. Surv. \textbf{26}~(4), 345--420 (1994).  doi:~\href{https://dx.doi.org/10.1145/197405.197406}{10.1145/197405.197406}.

\bibitem{DBLP:conf/pakdd/HuangZMLLWHD16en}
T.~Huang, Y.~Zhu, Y.~Mao et al., ``Parallel discord discovery,'' \textit{in Proc. of the 20th Pacific-Asia Conference on Advances in Knowledge Discovery and Data Mining, PAKDD 2016, Auckland, New Zealand, April~19--22, 2016}. 2016, pp.~233--244. DOI:~\href{https://dx.doi.org/10.1007/978-3-319-31750-2_19}{10.1007/978-3-319-31750-2\_19}.

\bibitem{KeoghLF05} 
E.~Keogh, J.~Lin, and A.~Fu, ``HOT SAX: efficiently finding the most unusual
time series subsequence,'' in \textit{Proc. of the 5th IEEE Int. Conf. on Data Mining, ICDM’05, Houston, Texas, November~27--30, 2005} (IEEE, 2005), pp.~8. doi:~\href{https://dx.doi.org/10.1109/ICDM.2005.79}{10.1109/ICDM.2005.79}.

\bibitem{DBLP:journals/kais/KeoghLLH07} 
E.~J.~Keogh, J.~Lin, S.~Lee, and H.~V.~Herle, ``Finding the most unusual time
series subsequence: algorithms and applications'',  Knowl. Inf. Syst. \textbf{11} (1), 1--27 (2007). DOI:~\href{https://doi.org/10.1007/s10115-006-0034-6}{10.1007/s10115-006-0034-6}.

\bibitem{DBLP:conf/dmkd/LinKLC03}
J.~Lin, E.~J.~Keogh, S.~Lonardi, and B.~Y.~Chiu, ``A symbolic representation of
time series, with implications for streaming algorithms,'' in \textit{Proc. of
the 8th ACM SIGMOD workshop on Research issues in data mining and
knowledge discovery, DMKD 2003, San Diego, California, USA, June~13,
2003} (ACM Press, 2003), pp.~2--11. doi:~\href{https://dx.doi.org/10.1145/882082.882086}{10.1145/882082.882086}.

\bibitem{DBLP:conf/sc/Mattson06en}
T.~Mattson, ``Introduction to OpenMP,'' in \textit{Proc. of the 2006 ACM/IEEE Conf. on Supercomputing, Tampa, FL, USA, November~11--17, 2006} (ACM Press, 2006), pp.~209.
doi:~\href{https://dx.doi.org/10.1145/1188455.1188673}{10.1145/1188455.1188673}.

\bibitem{Sodani15en}
A.~Sodani, ``Knights Landing (KNL): 2nd generation Intel Xeon Phi processor,'' in \textit{2015 IEEE Hot Chips 27th Symposium (HCS), Cupertino, CA, USA, August~22--25, 2015} (IEEE, 2015), pp.~1--24.	

\bibitem{DBLP:conf/hpcc/WuZHLLL15en}
Y.~Wu, Y.~Zhu, T.~Huang et al., ``Distributed Discord Discovery: Spark
based anomaly detection in time series,'' in \textit{Proc. of the 17th IEEE Int.
Conf. on High Performance Computing and Communications, HPCC 2015, 7th IEEE Int. Symp. on Cyberspace Safety and Security, CSS 2015, and 12th IEEE Int. Conf. on Embedded Software and Systems, ICESS 2015, New York, NY, USA, August~24--26, 2015} (IEEE, 2015), pp.~154--159. doi:~\href{https://dx.doi.org/10.1109/HPCC-CSS-ICESS.2015.228}{10.1109/HPCC-CSS-ICESS.2015.228}.

\end{biblio_lat}

\end{document}